\documentclass[12pt]{article}
\usepackage{latexsym, epsfig, verbatim,  amsmath, amssymb, lscape,multirow,color,enumitem}   %
\usepackage{threeparttable}
\usepackage{float}
\usepackage{graphicx}
\usepackage{subcaption}
\usepackage{natbib}

 \usepackage{amsmath}

\def\bfm#1{\mbox{\boldmath$#1$}}
{\addtolength{\baselineskip}{.3\baselineskip}}
\def\ba{\begin{array}}
\def\ea{\end{array}}
\def\be{\begin{equation}}
\def\ee{\end{equation}}

\newtheorem{theorem}{Theorem}

\newtheorem{proposition}{Proposition}  

\newcommand{\hy}{{\hat{y}}}

\newcommand{\bbI}{{\mathbb{I}}}

\newcommand{\bbR}{{\mathbb{R}}}
\newcommand{\bbS}{{\mathbb{S}}}
\newcommand{\bbT}{{\mathbb{T}}}

\newcommand{\bbY}{{\mathbb{Y}}}

\newcommand{\srf}[2]{\stackrel{{\rm #1}}{#2}}

\newcommand{\MC}[3]{\multicolumn{#1}{#2}{#3}}

\newcommand{\0}{{\bf 0}\!\!\!{\bf 0}}
        \def\b1{{\bf 1\!\!\!1}}  

\newcommand{\bp}{{\bf p}}
\newcommand{\bq}{{\bf q}}

\newcommand{\bu}{{\bf u}}

\newcommand{\bw}{{\bf w}}
\newcommand{\bx}{{\bf x}}
\newcommand{\by}{{\bf y}}

\newcommand{\bc}{{\bf c}}

\newcommand{\bA}{{\bf A}}
\newcommand{\bB}{{\bf B}}

\newcommand{\bI}{{\bf I}}

\newcommand{\bP}{{\bf P}}

\newcommand{\bR}{{\bf R}}

\newcommand{\bV}{{\bf V}}
\newcommand{\bW}{{\bf W}}
\newcommand{\bX}{{\bf X}}
\newcommand{\bY}{{\bf Y}}
\newcommand{\bZ}{{\bf Z}}

\newcommand{\bF}{{\bf F}}

\newcommand{\hg}{\hat{g}}
\newcommand{\hh}{\hat{h}}

\newcommand{\hbP}{\widehat{\bP}}

\newcommand{\hbq}{\widehat{\bq}}

\newcommand{\hth}{\hat{\theta}}

\newcommand{\hal}{\hat{\alpha}}   \newcommand{\hbe}{\hat{\beta}}

      \newcommand{\hbbe}{\widehat{{\bfm \beta}}}

\newcommand{\hbW}{\widehat{\bW}}

\newcommand{\hbw}{\widehat{\bw}}

\newcommand{\al}{\alpha}    

\newcommand{\bSi}{{\bfm \Sigma}}

\newcommand{\bthe}{{\bfm \theta}}

  \newcommand{\bbe}{{\bfm \beta}}

\newcommand{\bal}{{\bfm \alpha}}
\newcommand{\hbal}{{\widehat{\bal}}}
\newcommand{\bvarep}{{\bfm \varepsilon}}
\newcommand{\bga}{{\bfm \gamma}}
\newcommand{\hbga}{{\widehat{\bga}}}

\newcommand{\Cov}{\mbox{Cov}}          \newcommand{\cov}{\mbox{cov}} \newcommand{\var}{\mbox{var}}

\newcommand{\I}{\mbox{I}}

\renewcommand{\ge}{\geqslant}
\renewcommand{\le}{\leqslant}

\newcommand{\T}{\!\top\!}

\newcommand{\noi}{\noindent}

\def\bds{\begin{description} \itemsep=-\parsep \itemindent=-0.7 cm}
\def\eds{\end{description}}

\begin{document}

\title{Dynamic Tensor Recommender Systems}
\author{Yanqing Zhang,~~Xuan Bi,~~Niansheng Tang and~~Annie Qu\footnote{Correspondence to:
Dr. Annie Qu, Department of Statistics, University of Illinois at Urbana-Champaign, Champaign, IL 61820 (E-mail: anniequ@illinois.edu)
} \ \  \\
{\small Yunnan University, University of Minnesota and
University of Illinois at Urbana-Champaign
}  \\
}
\date{}
\maketitle
\vspace{1mm}

\begin{abstract}
Recommender systems have been extensively used by the entertainment industry, business marketing and the biomedical industry. In addition to its capacity of providing preference-based recommendations as an unsupervised learning methodology, it has been also proven useful in sales forecasting, product introduction and other production related businesses.
Since some consumers and companies need a recommendation or prediction for future budget, labor and supply chain coordination, dynamic recommender systems for precise forecasting have become extremely necessary. In this article, we propose a new recommendation method, namely the dynamic tensor recommender system (DTRS), which aims particularly at forecasting future recommendation. The proposed method utilizes a tensor-valued function of time to integrate time and contextual information, and creates a time-varying coefficient model for temporal tensor factorization through a polynomial spline approximation. Major advantages of the proposed method include competitive future recommendation predictions and effective prediction interval estimations.
 In theory, we establish the convergence rate of the proposed tensor factorization and asymptotic normality of the spline coefficient estimator. The proposed method is applied to simulations and IRI marketing data. Numerical studies demonstrate that the proposed method outperforms existing methods in terms of future time forecasting.

 \vskip 0.1in
\noi {\bf Keywords}:
 Contextual information,
Dynamic recommender systems,
Polynomial spline approximation,
Prediction interval,
Product sales forecasting
\end{abstract}

\section{Introduction}

Recommender systems (RS) are widely used in our daily lives, such as for selecting movies, restaurants, news articles, or online shopping.
As one of the information filtering techniques, RS can help users to find interesting items through combining several information sources, e.g., users' ratings and purchasing histories, item profiles and sales volumes, time, location, and companion or promotion strategies. Particularly, incorporating time is useful in RS since users' purchase behaviors are dynamic and often highly dependent on seasonal and time factors, and business sectors also rely on dynamic recommendations to track users' changing purchase interests over time. Thus, it is essential to capture information related to time and develop time-dependent RS, and we refer this as dynamic RS (DRS).

However, developing competitive DRS brings new challenges.
First, since data are streaming in over time and are time-dependent, general RS methods which are not capable of capturing time-dependency features may have reduced recommendation accuracy.
Second, forecasting future recommendations accurately is also a great challenge for DRS due to the complexity of changing users' interests. For example, users might like to watch news on weekdays, but watch movies on weekends. A shoe store sells more sandals in summer and more snow boots in winter.
It is important to borrow information from historical data in developing trends. Many RS methods are not designed to capture trends and
  predict future recommendations. In addition, as data are streaming in over time, future recommendations could involve new users or new items, whose information is not available from historical data. This is also a common problem encountered in RS, referred as the ``cold start'' problem.

General RS approaches include content-based filtering and collaborative filtering (CF).
Traditionally, content-based filtering methods recommend similar types of items by matching a user's preferred item profile with current item's profile \citep[e.g.,][]{Salter2006,SON2017404}.
In contrast, CF methods recommend items by predicting item ratings for the active user based on ratings from other similar users \citep[e.g.,][]{Herlocker2004,LUO2012271}.
On the basis of CF methods, research work related to DRS have been developed in recent years \citep[e.g.,][]{Koren2009,Gultekin2014,Yu2016,Wu:2017,GUO201856,xiong2010temporal,Rafailidis2014MDU,bi2018,Wu2019NTF}.
However, most of these methods can only make recommendations for observed discrete time points, and are not designed for future recommendation prediction on unobserved time points;
for example, matrix factorization incorporating periodic and continual temporal effects \citep{GUO201856},
coupled tensor factorization exploiting users' demographic information \citep{Rafailidis2014MDU} and
the collaborative Kalman filter \citep{Gultekin2014}.
CF methods incorporating a time series model \citep{Yu2016} or incorporating long short-term memory modeling \citep{Wu:2017,Wu2019NTF} are able to solve the forecasting problem,
but cannot deal with new users, items or contextual variables.
\cite{xiong2010temporal} used a
 Bayesian estimation procedure with a time-dependent constraint to predict DRS for new users and items,
  while \cite{bi2018} created an additional layer of nested latent factors for new time points, users and items. However, both methods require discrete  time points for constructing a tensor.

In this article, we propose a new time-varying coefficient model for the DRS based on tensor canonical polyadic decomposition (CPD); namely, the dynamic tensor recommender system (DTRS).
Specifically, we introduce a tensor-valued function of time with each mode corresponding to \emph{user}, \emph{item} or a \emph{contextual variable}, where each component of the tensor is a function of time and has intra-cluster correlation. In the CPD framework, we build a time-varying coefficient model incorporating group information of time points, users, items and contexts.
We approximate each coefficient function by a polynomial spline and employ group factors to explore homogeneous group effects.
We adopt the weighted least square approach to incorporate intra-cluster correlation for more efficient estimation.
In addition, we construct the prediction intervals of estimators of tensor components to forecast the confidence range of predicted values.
In theory, we establish
the convergence rate of the proposed tensor factorization and the
asymptotic property of the spline parametric estimator.

The proposed method has two significant contributions.
First, it can effectively forecast recommendations at future time points. This is because the proposed model integrates time dependency to the DRS through time-varying coefficient modeling in tensor factorization so that it can effectively capture dynamic trends of DRS. In addition, the subgroup factors in the proposed model extract homogeneous information from the same group, which provides recommendation forecasting for future time points and therefore solves the ``cold start'' problem. In contrast to general CF methods which require discrete time points as a tensor mode, the proposed approach is more flexible by utilizing a continuous tensor-value function.

Second, the proposed method is able to provide pointwise prediction intervals. In practice, it is desirable to know the upper and lower bound for predictions, such as the highest possible cost, or the future sales volumes or revenues in the worst case scenario. However, existing methods on prediction intervals are mostly univariate or multivariate time series, and the prediction intervals for user-item-context interactions in a tensor framework have not been developed. The proposed approach develops the prediction intervals for component estimators of a tensor-valued function, which provide a more complete picture of the DRS over time. In our real data analysis, the proposed approach provides effective prediction interval estimators of the sales volumes for IRI marketing data \citep{Bronnenberg2008}, which can help store managers to make sound decisions on marketing strategy and inventory planning.

The remainder of the paper is organized as follows.
Section 2 introduces the notation and background on tensor and tensor factorization.
Section 3 presents the proposed method and its implementation.
Theoretical properties are derived in Section 4.
Section 5 presents simulation studies to assess the performance of the proposed approach.
In Section 6, we apply the proposed method to the IRI marketing data.
 Concluding remarks and discussion are provided in Section 7.

\section{Notation and Background}

In this section, we introduce the background of the tensor and some notation.
Throughout this article, we use blackboard capital letters for sets, e.g., $\bbT,\bbI$, small letters for scalars, e.g., $x,y\in\bbR$,
bold small letters for vectors, e.g., $\bx, \by\in\bbR^n$, bold capital letters for matrices, e.g., $\bX, \bY\in\bbR^{n_1\times n_2}$,
and Euler script fonts for tensors, e.g., $\mathcal{X},\mathcal{Y}\in\bbR^{n_1\times n_2\times\cdots\times n_d}\; (d>2)$.

A $d$th-order tensor is an array with $d$ dimensions ($d>2$), which is an extension of a matrix to higher order. Here $d$ represents the tensor's order.
We denote the component $(i_1,i_2,\cdots, i_d)$ of a $d$th-order tensor $\mathcal{Y}$ by $y_{i_1i_2\cdots i_d}$, where $i_k=1,2,\ldots,n_k$, and $k$ is called a mode of the tensor ($k=1,2,\ldots,d$ ).
In particular, a tensor $\mathcal{Y}$ is called a rank-one tensor if it can be written as
$\mathcal{Y}=\bp^1\circ\bp^2\circ\cdots\bp^d$,
where the symbol $\circ$ represents the vector outer product, and
$\bp^k$ is a $n_k$-dimensional latent factor corresponding to the $k$th mode.
That is, each component of the tensor is the product of the corresponding
vector components: $y_{i_1i_2\cdots i_d}=p^1_{i_1}p^2_{i_2}\cdots p^d_{i_d}$.

\begin{figure}[H]
\centering
\scalebox{0.5}[0.5]{\includegraphics{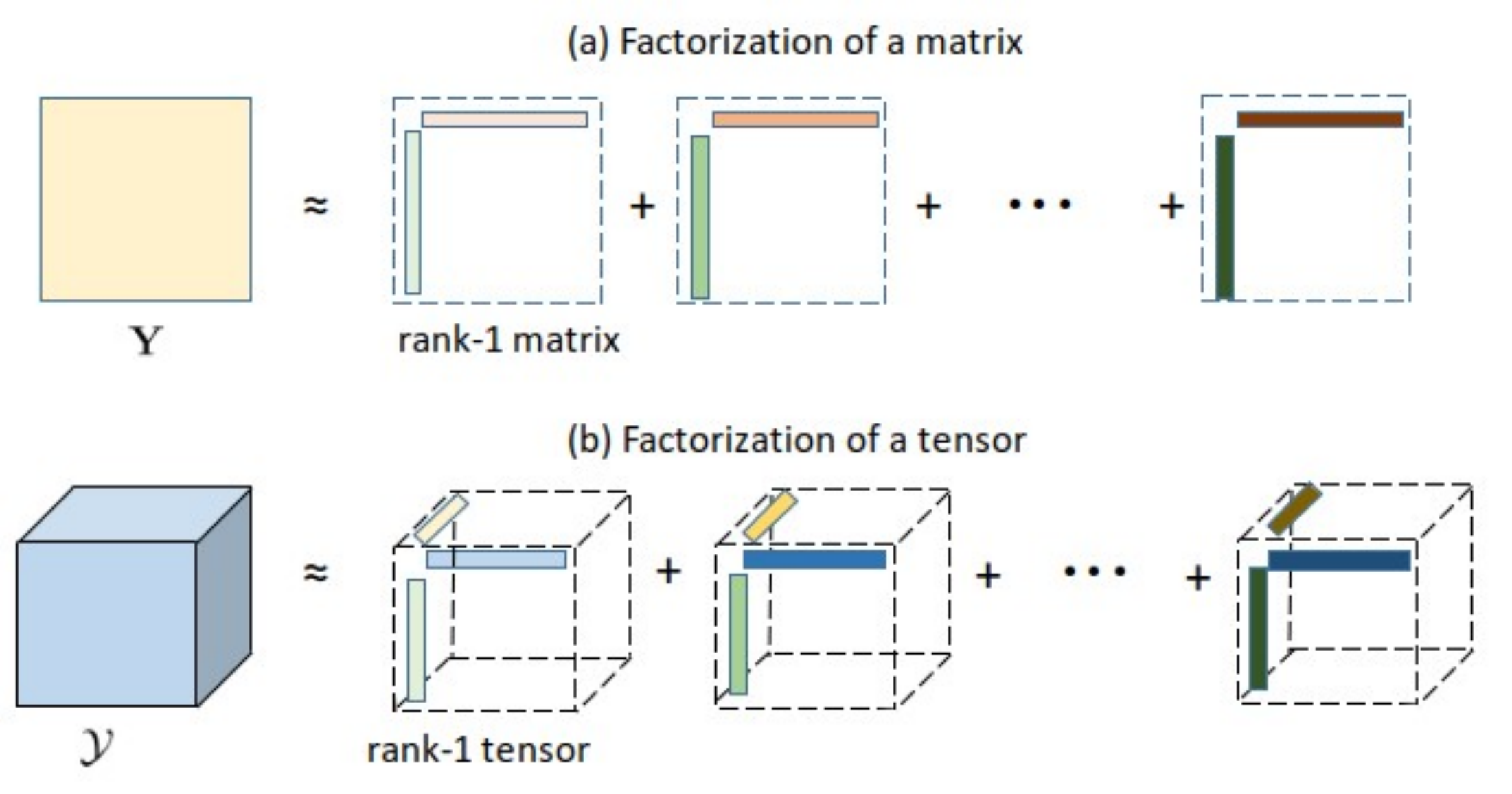}}\vspace{-3mm}
\caption{\footnotesize{Illustration of factorizations of a matrix and a third-order tensor. (a) factorization of a matrix into $r$ rank-1 matrices, (b) CPD of a third-order tensor into $r$ rank-1 tensors.}}
\label{factorization}
\end{figure}

 The canonical polyadic decomposition (CPD) is commonly adopted in tensor decomposition, which decomposes a tensor as a sum of $r$ rank-one
tensors. That is:
$$
\mathcal{Y}\approx\sum^r_{j=1}\bp^1_{\cdot j}\circ\bp^2_{\cdot j}\circ\cdots\circ\bp^d_{\cdot j},
$$
where $\bp^k_{\cdot j}=(p^k_{1j},\cdots,p^k_{n_kj})^{\T}$ is a $n_k$-dimensional latent factor corresponding to the $k$th mode for $k=1,\ldots,d; j=1,\ldots,r$.
 Equivalently, each component of $\mathcal{Y}$ is
$$
y_{i_1i_2\cdots i_d} \approx\sum^r_{j=1}p^1_{i_1j}p^2_{i_2j}\cdots p^d_{i_dj}.
$$
The CPD can be considered to be a higher-order generalization of matrix factorisation. Figure \ref{factorization} illustrates a matrix factorization of a matrix and a CPD of a third-order tensor.
An extensive review of tensors and other forms of tensor decomposition are discussed in \cite{Kolda2009Bader}.

Let $\bP^k=(\bp^k_{\cdot1},\bp^k_{\cdot2},\ldots,\bp^k_{\cdot r})_{n_k\times r}$ and $\bthe=\{\bP^1,\bP^2,\cdots,\bP^d\}$. We can estimate $\bthe$ via minimizing a loss function (e.g., $L_2$ loss). However, the non-convexity of the loss function could impose computational complexity due to numerical instability or even non-convergence \citep{Silva2008Lim,Frolov2017}.
A common approach to alleviate the non-convexity problem is to
introduce regularization. We define an objective function with a penalty function as the following:
$$
L(\bthe|\mathcal{Y})=Q(\mathcal{Y},\bthe)+J(\bthe),
$$
where $Q$ is a loss function and $J$ is a penalty function, such as $L_2$, $L_1$ or $L_0$ penalties, or a fused Lasso.

Specially, the optimization problem solves
$\bthe^*=\arg\min L(\bthe|\mathcal{Y})$,
where $\bthe^*$ defines an optimal set of model parameters.
In the case of squared loss function with an $L_2$-penalty, the objective function
is
$$
L(\bthe|\mathcal{Y})=\sum_{(i_1,i_2,\ldots,i_d)\in\Omega}(y_{i_1i_2\cdots i_d}-\sum^r_{j=1}p^1_{i_1j}p^2_{i_2j}\cdots p^d_{i_dj})^2+\lambda\sum^d_{k=1}\|\bP^k\|^2_F,
$$
where $\|\cdot\|_{F}$ represents the Frobenius norm, and $\Omega=\{(i_1,i_2,\ldots,i_d): y_{i_1i_2\ldots i_d}\;\; {\rm is\;\; observed}\}$
is a set of indices corresponding to the observed components.
Notice that, in the context of RS, the set $\Omega$ may not contain all indices of the tensor components and could be a small fraction of the entire tensor size, since the majority of the tensor components could be missing.
 Major algorithms for implementing the optimization problem include the cyclic coordinate descent algorithm, the stochastic gradient descent method and the maximum block improvement algorithm \citep{Chen2012He}.

\section{The Proposed Method}

\subsection{General Methodology}

In this subsection, we develop the methodology for the proposed DTRS method. Specifically, we adopt the idea of a time-varying coefficient model under the CPD framework to capture the trends of the DRS, and classify time points into subgroups to infer new time point trends through existing time points of the same group.

\begin{figure} [H]
\centering
\scalebox{0.5}[0.4]{\includegraphics{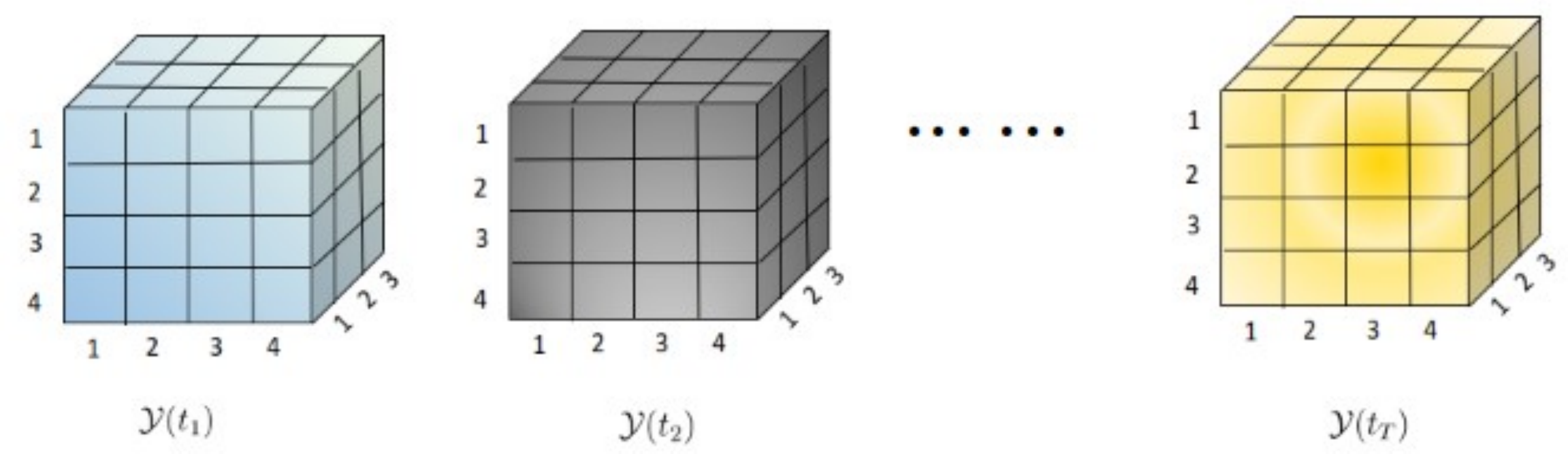}} \vspace{-3mm}
\caption{\footnotesize{Third-order tensor-valued process.}}
\label{yt}
\end{figure}

 We consider a $d$th-order tensor-valued function $\mathcal{Y}(t)\in\bbR^{n_1\times n_1\times\ldots\times n_d}$, where the value at time $t$ is a $d$-dimensional array. The tensor set $\bbY=\{\mathcal{Y}(t_i): t_i\in \bbT, i=1,2,\ldots,T\}$ is the corresponding stochastic process defined on a compact interval $\bbT$.  Without loss of generality, let $\bbT$ be a closed interval $[0, 1]$. Notice that we do not require $t_i$'s to be equally distanced as in other time-dependent tensor models \citep[e.g.,][]{xiong2010temporal,bi2018}, so that the proposed method can be applied for a recommender system with arbitrary time points. Figure \ref{yt} illustrates an example of a tensor-valued process with $d=3$. In the DRS, the tensor-valued process could be the rating or sale volume of items or products from users or stores given contexts.
We assume that time points can be categorized into different subgroups, where time points of the same group have common information. For example, in our numerical studies, time points in the same month from the twelve months of each year are categorized in the same group. In addition to time, we also categorize subjects from other modes into subgroups if they share similar characteristics, for example, stores of the same market and products of the same product category.

Suppose the subgroup labels are given, we formulate each component of $\mathcal{Y}(t)$ as follows:
\begin{equation}\label{cpd2}
y_{i_1i_2\cdots i_d}(t)=\sum^r_{j=1}h_j(t)p^1_{i_1j}p^2_{i_2j}\cdots p^d_{i_dj}+g(t)q^1_{i_1}q^2_{i_2}\cdots q^d_{i_d}+\varepsilon_{i_1i_2\ldots i_d}(t),
\end{equation}
where $\varepsilon_{i_1i_2\ldots i_d}(t)$ is a stochastic process with mean zero and finite variance,
 $h_j(t)$ is a trend function of time,
$p^k_{i_kj}$ and $q^k_{i_k}$ are the $j$th latent factor and the subgroup factor for the $i_k$th subject from the $k$th mode, respectively,
$k=1,2,\ldots d$, $j=1,2,\ldots,r$,
and $g(t)=\sum_{e=1}^{m_{d+1}}g_e(t)I(t\in s_e)$, in which $I(\cdot)$ is an indicator function, $m_{d+1}$ is the number of subgroups for time, and $g_e(t)$ is a trend function corresponding to the $e$th subgroup $s_e$ of time.
We have $q^k_{i_k}=q^k_{i'_k}=q^k_{(e_k)}$ if the $i_k$th and $i'_k$th subjects are from the $e_k$th subgroup ($e_k=1,2,\ldots,m_k$), where $q^k_{(e_k)}$ is the subgroup factor associated with the $e_k$th subgroup, and $m_k$ is the number of subgroups for the $k$th mode, $k=1,2,\ldots,d$. We denote the set of observed time points for  the component $y_{i_1i_2\cdots i_d}(t)$ by $\bbT_{i_1i_2\ldots i_d}$, and the number of components of this set by $|\bbT_{i_1i_2\ldots i_d}|$. Let $\by_{i_1i_2\ldots i_d}=\{y_{i_1i_2\cdots i_d}(t)\}_{t\in\bbT_{i_1i_2\ldots i_d}}$ and $\bvarep_{i_1i_2\ldots i_d}=\{\varepsilon_{i_1i_2\ldots i_d}(t)\}_{t\in\bbT_{i_1i_2\ldots i_d}}$ be $|\bbT_{i_1i_2\ldots i_d}|\times1$ vectors. We assume that the covariance matrix is $\cov(\by_{i_1i_2\ldots i_d})=\cov(\bvarep_{i_1i_2\ldots i_d})=\bSi^0_{i_1i_2\ldots i_d}$,
typically not an identity matrix due to the intra-cluster correlation arising from repeated observed data.

 Model (\ref{cpd2}) adopts the idea of varying-coefficient models to create a CPD for tensor data. Varying-coefficient models are a useful tool to explore dynamic patterns, and have been applied to modeling and predicting longitudinal, functional, and time series data \citep{Huang2004Shen,Fan2008Zhang}. The first part of equation (\ref{cpd2}) is an individual-level factor model which takes into account the heterogeneity of subjects and trend of time, and the time-varying coefficients $h_j(t)$ ($j=1,\ldots,r$) reflect the dynamic features.
The second part of equation (\ref{cpd2}) is a subgroup-level factor model to capture common features from the same subgroups, where the subgroup factors can accommodate new subjects from any mode at future time points, and the $g(t)$ allows time variables to follow a subgroup function of time such that we can predict future time points via borrowing information from existing time points of the same group.

To capture these trend functions, we adopt the polynomial splines to approximate $h_j(t)$ and $g_e(t)$.
Let $\{\nu_{ji}\}_{i=1}^{a_N}$ be interior knots within $\bbT$, and $\Upsilon_j$ be a partition of $\bbT$ with $a_N$ knots, that is $\Upsilon_j=\{0=\nu_{j0}<\nu_{j1}<\cdots<\nu_{ja_N}<\nu_{ja_N+1}=1\}$ for $j=1,2,\ldots,d$. The polynomial splines of an order $\kappa+1$ are functions with $\kappa$-degree of polynomials on intervals $[\nu_{ji-1},\nu_{ji})$ for $i=1,2,\ldots,a_N$ and $[\nu_{ja_N},\nu_{ja_N+1}]$, and have $\kappa-1$ continuous derivatives globally. Denote a spline bases vector of the space of such spline functions as $\bB_j(t)=(B_{j1}(t),\ldots,B_{jM}(t))^{\T}$, where $M=a_N+\kappa+1$ as the number of spline bases.
 The function $h_j(t)$ ($j=1,2,\ldots,d$) can be approximated by
$$
\begin{array}{llll}
\hat{h}_j(t) & = & \sum_{i=1}^M\al_{ji}B_{ji}(t)=\bal_j^{\T}\bB_j(t),
\end{array}
$$
where $\bal_j=(\al_{j1},\al_{j2},\ldots,\al_{jM})^{\T}$ is a coefficient vector.
Spline functions can be B-spline or truncated polynomial functions. For example, for the truncated polynomial function,
$\bB_j(t)=(1, t, \ldots, t^{\kappa}, (t-\nu_{j1})_+^{\kappa},\ldots,(t-\nu_{ja_N})_+^{\kappa})^{\T}$, and the $(t-
\nu)_+$ is $t-\nu$ if $t>\nu$ and 0 otherwise.

Similarly, let $\{\omega_{ei}\}_{i=1}^{a_N}$ be interior knots within $\bbT$,
$\Gamma_e=\{0=\omega_{e0}<\omega_{e1}<\cdots<\omega_{ea_N}<\omega_{ea_N+1}=1\}$,
and $\bA_e(t)=(A_{e1}(t),\ldots,A_{eM}(t))^{\T}$ be a vector of spline bases
for $e=1,2,\ldots,m_{d+1}$.
The $g_e(t)$ can be approximated by
$$
\hat{g}_e(t)=\sum_{i=1}^M\beta_{ei}A_{ei}(t)=\bbe_e^{\T}\bA_e(t),
$$
where $\bbe_e=(\beta_{e1},\beta_{e2},\ldots,\beta_{eM})^{\T}$.
Thus, the prediction based on equation (\ref{cpd2}) is
\begin{equation}\label{cpd3}
\hy_{i_1i_2\cdots i_d}(t)=\sum^r_{j=1}\hh_j(t)p^1_{i_1j}p^2_{i_2j}\cdots p^d_{i_dj}+\hg(t)q^1_{i_1}q^2_{i_2}\cdots q^d_{i_d},
\end{equation}
where $\hg(t)=\sum_{e=1}^{m_{d+1}}\hg_e(t)I(t\in s_e)$.
The model (\ref{cpd3}) can capture trends of the DRS sufficiently through the polynomial spline approximations of time-varying coefficient functions.
In addition, since the spline approximation is computationally fast \citep{Xue2006Yang}, the model (\ref{cpd3}) can achieve the spline estimates of the coefficients efficiently, and this is especially advantageous in estimating high-dimensional parameters in RS.

Due to the intra-cluster correlation, it is important to incorporate intra-cluster correlation into RS. However, in practice, the covariance matrix $\bSi_{i_1i_2\ldots i_d}^0$ is often unknown. We adopt an invertible working covariance matrix, denoted as $\bSi_{i_1i_2\ldots i_d}$, to take into account the intra-cluster correlation.
Let
$\bP=(\bP^{1\T},\cdots,\bP^{d\T})$,
$\bq=(\bq^{(1)\T},\cdots,\bq^{(d)\T})^{\T}$,
$\bal=(\bal_1^{\T},\ldots,\bal_r^{\T})^{\T}$,
$\bbe=(\bbe_1^{\T},\ldots,\bbe_{m_{d+1}}^{\T})^{\T}$,
and $\bga=(\bal^{\T},\bbe^{\T})^{\T}$,
where
$\bP^k=(\bp^k_{\cdot1},\ldots,\bp^k_{\cdot r})$, $\bp^k_{\cdot j}=(p^k_{1j},\cdots,p^k_{n_kj})^{\T}$, $\bq^{(k)}=(q^k_{(1)},\ldots,q^k_{(m_k)})^{\T}$,
and $k=1,\ldots,d$. Define $\bthe=\{\bP,\bq,\bga\}$ as parameters of interest.  
We define the following weighted penalized objective function:
\begin{equation}\label{obj}
L(\bthe|\bbY)=\sum_{(i_1,i_2,\cdots,i_d)\in\Omega}(\by_{i_1i_2\ldots i_d}-\widehat{\by}_{i_1i_2\ldots i_d})^{\T}\bSi_{i_1i_2\ldots i_d}^{-1}
(\by_{i_1i_2\ldots i_d}-\widehat{\by}_{i_1i_2\ldots i_d})+\lambda(\|\bP\|^2_F+\|\bq\|^2_2+\|\bga\|^2_2),
\end{equation}
where $\lambda$ is the penalized parameter, $\Omega=\{(i_1,i_2,\ldots,i_d): y_{i_1i_2\cdots i_d}(t)\; {\rm is \; observed\; at\; some\;} t\}$,
$N=|\Omega|$ is the number of components of the set $\Omega$,
$\|\cdot\|_2$ is the Euclidean norm, and
$\widehat{\by}_{i_1i_2\ldots i_d}=\{\hy_{i_1i_2\cdots i_d}(t)\}_{t\in\bbT_{i_1i_2\ldots i_d}}$ is a $|\bbT_{i_1i_2\ldots i_d}|\times1$ vector.

The matrix $\bSi_{i_1i_2\ldots i_d}$
is an approximation of the true covariance $\bSi_{i_1i_2\ldots i_d}^0$, and can be modeled as $\bSi_{i_1i_2\ldots i_d}=\bV_{i_1i_2\ldots i_d}^{1/2}\bR_{i_1i_2\ldots i_d}\bV_{i_1i_2\ldots i_d}^{1/2}$, where
$\bV_{i_1i_2\ldots i_d}$ is a diagonal matrix of the marginal variance of $\by_{i_1i_2\ldots i_d}$, and $\bR_{i_1i_2\ldots i_d}$ is a working correlation matrix for $\by_{i_1i_2\ldots i_d}$.
Some commonly used working correlation structures include independence, exchangeable, and first-order autoregressive process (AR-1), among others.
Given a working correlation structure, the working correlation matrix depends on fewer nuisance parameters which can be estimated by the residual-based moment method \citep{Liang1986Zeger}. The proposed method is robust to the misspecification
of correlation structure as indicated by our numerical examples.

\subsection{Parameter Estimation}

In this subsection, we discuss parameter estimation by minimizing (\ref{obj}).
Let $\bp^k_{i_k}=(p^k_{i_k1},\cdots,\\ p^k_{i_kr})^{\T}$ and
$\Omega^k_{i_k}=\{(i_1,\ldots,i_k,\ldots,i_d): y_{i_1\cdots i_d}(t)\; {\rm is \; observed\; at\; some\;} t {\rm\; given\;} i_k\}$
be the set of indices with the fixed $k$th mode index $i_k$, where the corresponding components are observed at some time points. We assume that the number of observations for each time subgroup $s_e$ is larger or equal than 2 for $e=1,\ldots,m_{d+1}$, and the number of observations for each subgroup $e_k$ from the $k$th mode is larger or equal than 2 for $e_k=1,\ldots,m_k; k=1,\ldots, d$.
The partial derivatives of $L(\cdot|\bbY)$ have explicit forms with respect to the individual factors, the subgroup factors and the spline coefficients, which makes it feasible to apply the blockwise coordinate descent approach (BCD). That is,
for $i_k=1,\ldots,n_k$ and $k=1,\ldots,d$,
\begin{equation}\label{objf1}
\widehat{\bp}^k_{i_k} =\arg\min_{\bp^k_{i_k}}\sum_{\Omega^k_{i_k}}(\by_{i_1i_2\ldots i_d} - \widehat{\by}_{i_1i_2\ldots i_d})^{\T}\bSi_{i_1i_2\ldots i_d}^{-1}
(\by_{i_1i_2\ldots i_d} - \widehat{\by}_{i_1i_2\ldots i_d})+\lambda\|\bp^k_{i_k}\|^2_2,
\end{equation}
\begin{equation}\label{objf2}
\widehat{\bq}^{(k)} =\arg\min_{\bq^{(k)}}\sum_{\Omega}(\by_{i_1i_2\ldots i_d} - \widehat{\by}_{i_1i_2\ldots i_d})^{\T}\bSi_{i_1i_2\ldots i_d}^{-1}
(\by_{i_1i_2\ldots i_d} - \widehat{\by}_{i_1i_2\ldots i_d})+\lambda\|\bq^{(k)}\|^2_2,
\end{equation}
\begin{equation}\label{objf3}
\hbal =\arg\min_{\bal}\sum_{\Omega}(\by_{i_1i_2\ldots i_d} - \widehat{\by}_{i_1i_2\ldots i_d})^{\T}\bSi_{i_1i_2\ldots i_d}^{-1}
(\by_{i_1i_2\ldots i_d} - \widehat{\by}_{i_1i_2\ldots i_d})+\lambda\|\bal\|^2_2,
\end{equation}
\begin{equation}\label{objf4}
\hbbe =\arg\min_{\bbe}\sum_{\Omega}(\by_{i_1i_2\ldots i_d} - \widehat{\by}_{i_1i_2\ldots i_d})^{\T}\bSi_{i_1i_2\ldots i_d}^{-1}
(\by_{i_1i_2\ldots i_d} - \widehat{\by}_{i_1i_2\ldots i_d})+\lambda\|\bbe\|^2_2.
\end{equation}

In fact, the estimation procedure of $\widehat{\bp}^k_{i_k}$ in (\ref{objf1}) is a ridge regression, and does not require knowing $\bp^k_{i'_k}$ for $i'_k\neq i_k$.
Thus, parallel computation is applicable to calculate $\widehat{\bp}^k_1,\ldots,\widehat{\bp}^k_{n_k}$ efficiently.
The minimization of $L(\bthe|\bbY)$ can be done cyclically through estimating $\bP$, $\bq$, $\bal$ and $\bbe$.
Notice that $\Omega=\cup^{n_k}_{i_k=1}\Omega^k_{i_k}$, and it is possible that $\Omega^k_{i_k}$ is empty for certain $i_k$'s, that is, there is no observation on the subject $i_k$. Under this circumstance,
the individual factor of the $i_k$ subject is assigned as $\bp^k_{i_k}=\0$, and the predicted values may degenerate to the subgroup-level factor model by utilizing information from members of the same subgroup.

\subsection{Implementation}

In the following, we discuss several implementation issues. To solve the objective function (\ref{obj}), we incorporate the maximum block improvement (MBI) strategy \citep{Chen2012He} into the BCD algorithm cyclically as in \cite{bi2018}. The MBI has two advantages over traditional cyclic BCD algorithms. First, it has a good algorithmic property which guarantees convergence to a stationary point, whereas traditional BCDs may end up with certain points where the criterion function ceases to decrease \citep{Chen2012He}. Second, the MBI has the capability of choosing descending directions and hence has the possibility to discover ``shortcuts'', which may reduce the computational time significantly.
Let $\widehat{\bthe}_{l}$ be an estimator of $\bthe$ at the $l$th iteration,
$\bthe_{a}$ be a subset of $\bthe$, $\bthe^{c}$ be the complementary set of $\bthe_{a}$, and $\widehat{\bthe}^{*}_a$ be the attempted update of $\bthe_{a}$. The improvement of the $\widehat{\bthe}^{*}_a$ is defined as
\begin{equation}\label{imp}
J_{\hth^*_a} = 1-\frac{L(\widehat{\bthe}^{*}_a,\widehat{\bthe}_{l-1}^c|\bbY)}{L(\widehat{\bthe}_{l-1}|\bbY)}.
\end{equation}
We summarize the implementation of the specifical algorithm as follows.

\newpage
\noindent $\overline{\mbox{\underline{\makebox[\textwidth]{\textbf{Implementation Algorithm}}}}}$
\begin{itemize}
\item[1.] (Initialization) Input all observed $y_{i_1i_2\cdots i_d}(t)$'s, the number of factors $r$, tuning parameter $\lambda$, initial value
$\bthe_0$ and a stopping criterion $\varepsilon=10^{-4}$.
\item[2.] (Individual factors update) At the $l$th iteration, estimate $\{\bP^1,\bP^2,\cdots,\bP^d,\bal\}$.
\begin{itemize}
\item[(i)] For each $\bP^k$, solve (\ref{objf1}) through parallel computing and obtain
$\widehat{\bP}^{k*}$. Then calculate $J_{\hbP^{k*}}$ through (\ref{imp}).
\item[(ii)] For $\bal$, solve (\ref{objf3}) and obtain
$\widehat{\bal}^*$. Then calculate $J_{\hal^*}$ through (\ref{imp}).
\item[(iii)] Assign \\
$\widehat{\bP}_l^{k}\leftarrow \widehat{\bP}^{k*}$, if $J_{\hbP^{k*}}=\max\{J_{\hbP^{1*}},J_{\hbP^{2*}},\cdots,J_{\hbP^{d*}},J_{\hal^{*}}\}$. \\
$\widehat{\bal}_{(l)}\leftarrow \widehat{\bal}^*$, if
$J_{\hal^{*}}=\max\{J_{\hbP^{1*}},J_{\hbP^{2*}},\cdots,J_{\hbP^{d*}},J_{\hal^{*}}\}$.
\end{itemize}

\item[3.] (Subgroup factors update) At the $l$th iteration, estimate $\{\bq^{(1)},\bq^{(2)},\cdots,\bq^{(d)},\bbe\}$.
    \begin{itemize}
    \item[(i)] For every $\bq^{(k)}$, solve (\ref{objf2}) and obtain $\hbq^{(k)*}$.
    Then calculate $J_{\hbq^{(k)*}}$ through (\ref{imp}).
    \item[(ii)] For $\bbe$, solve (\ref{objf4}) and obtain
    $\widehat{\bbe}^*$. Then calculate $J_{\hbe^*}$ through (\ref{imp}).
    \item[(iii)] Assign \\
    $\hbq_l^{(k)}\leftarrow \hbq^{(k)*}$, if $J_{\hbq^{(k)*}}=\max\{J_{\hbq^{(1)*}},J_{\hbq^{(2)*}},\cdots,J_{\hbq^{(d)*}},J_{\hbe^*}\}$. \\
    $\hbbe_{(l)}\leftarrow \hbbe^*$, if $J_{\hbe^*}=\max\{J_{\hbq^{(1)*}},J_{\hbq^{(2)*}},\cdots,J_{\hbq^{(d)*}},J_{\hbe^*}\}$.
    \end{itemize}
\item[4.] (Stopping Criterion) Stop if
$\max\{J_{\hbP^{1*}},J_{\hbP^{2*}},\cdots,J_{\hbP^{d*}},J_{\hal^*},J_{\hbq^{(1)*}},\cdots,J_{\hbq^{(d)*}},J_{\hbe^*}\}<\varepsilon$.
Set the final estimator $\widehat{\bthe}=\widehat{\bthe}_{l}$. Otherwise set $l\leftarrow l+1$
and go to step 2.
\end{itemize}
\noindent\makebox[\linewidth]{\rule{\textwidth}{0.4pt}}

To select tuning parameter $\lambda$, we search the one from grid points minimizing the root mean square
error on the validation set, defined as
$[\sum_{(i_1,\ldots,i_d,t)\in\Gamma}\{y_{i_1\ldots i_d}(t)-\hy_{i_1\ldots i_d}(t)\}^2/|\Gamma|]^{1/2}$,
 where $\Gamma$ is the set of indices and times of observed data.
We choose the number of individual latent factors $r$ such that it is sufficiently large and leads to stable estimation. In general,
the $r$ is no smaller than the theoretical rank of the tensor in order to
represent subjects' latent features sufficiently well, but not so large as to over-burden  the computational cost.

An appropriate selection of the knot sequence is important to efficiently implement the proposed method.
In practice, knot locations are usually chosen to be equally-spaced over the range of data or placed at evenly-spaced quantiles of data.
Since there are high-dimensional factor parameters, for simplicity we set the number of knots to be the integer part of $N^{1/(2\kappa+3)}$, where $N=|\Omega|$ and $\kappa$ is the degree of polynomials. One can also choose other methods to select the number of knots such as the AIC or BIC procedures
\citep{Xue2006Yang}.
The degree of polynomials $\kappa$ is commonly chosen as $1,2,$ or $3$.
In our numerical study, we set $\kappa=2$ and adopt truncated polynomial bases.
One can also use different degrees and spline bases for different time-varying coefficients.

Another important issue is the selection of contextual variables as tensor modes. On the one hand, a higher-order tensor with more contextual variables allows higher-order interactions and hence provides more accurate estimation. On the other hand, a higher-order tensor entails more complex and intensive computation, and may lead to overfitting. Thus, it is still an important open problem to determine which contextual variables should
be included in the tensor. In our numerical studies, promotion strategies are incorporated as a contextual variable, since users' and items' behaviors are distinctive under different promotion strategies. In general practice, however, we assume that the order of a tensor can be determined based on prior knowledge.

\section{Theoretical Properties}
\label{theoremsection}

In this section, we derive asymptotic properties for the proposed method. Specifically, we establish the convergence rate of the proposed tensor
factorization and the asymptotic normality of the spline coefficient estimator. Note that identifiability is critical for tensor representation.
We first present the sufficient conditions to ensure identifiability of the proposed tensor modeling as follows.
\begin{proposition} \label{pro1}
If $\sum_{k=1}^dK_{k}\ge2r+d+1$ holds, minimizers of $L(\bP,\bq,\bal,\bbe|\bbY)$ in $\bP$, $\bq$, $\bal$ and $\bbe$ given fixed spline bases are unique up to permutation almost surely, where $K_{k}$ is the Kruskal rank of $(\bP^k,\bq^k)$, and $\bq^k=(q^k_1,q^k_2,\cdots,q^k_{n_k})^{\T}$.
\end{proposition}

Proposition \ref{pro1} shows that the proposed tensor modeling is identifiable up to permutation almost surely.
To address permutation indeterminacy,
we could align the factors according to a descending order of the first row of mode-1
factor matrix $\bP^1$, that is, $p^1_{11}\ge p^1_{12}\ge\cdots\ge p^1_{1r}$, following the method in \cite{Zhang2014Wang}.
The rearrangement can be implemented during or after the proposed algorithm, since it does not affect the estimation procedure.
In the rest of Section 4, we assume that the parameters are identifiable.

Let
$\bfm{u}_{i_1i_2\ldots i_d}=\{(p^1_{i_11}p^2_{i_21}\cdots p^d_{i_d1}), (p^1_{i_12}p^2_{i_22,}\cdots p^d_{i_d2}),\cdots,
(p^1_{i_1r}p^2_{i_2r}\cdots p^d_{i_dr}),(q^1_{i_1}q^2_{i_2}\cdots q^d_{i_d})\}^{\T}$,
$\mathcal{U}\in\bbR^{n_1\times\ldots\times n_d\times (r+1)}$ consist of $\bfm{u}_{i_1i_2\ldots i_d}$,
${\bf f}(t)=\{h_1(t),h_2(t),\ldots,h_r(t),g(t)\}^{\T}$,
$\bF_{i_1i_2\ldots i_d}\in\bbR^{|\bbT_{i_1i_2\ldots i_d}|\times(r+1)}$ be the matrix consisting of ${\bf f}(t)$ for all $t\in\bbT_{i_1i_2\cdots i_d}$.
We rewrite the equation (\ref{cpd2}) as
$y_{i_1i_2\cdots i_d}(t)={\bf f}(t)^{\T}\bfm{u}_{i_1i_2\ldots i_d}+\varepsilon_{i_1i_2\ldots i_d}(t)$ for $t\in\bbT_{i_1i_2\cdots i_d}$. Thus, the corresponding vector form is
$$
\by_{i_1i_2\ldots i_d}=\bF_{i_1i_2\ldots i_d}\bfm{u}_{i_1i_2\ldots i_d}+\bvarep_{i_1i_2\ldots i_d}.
$$
Let $J(\mathcal{U})$ be a non-negative penalty function of $\mathcal{U}$.
The overall criterion given $h_j(\cdot)$ and $g(\cdot)$ is redefined as
 \begin{equation}\label{objU}
 L(\mathcal{U}|\bbY)=\sum_{(i_1,i_2,\cdots ,i_d)\in\Omega}(\by_{i_1i_2\ldots i_d} - \bF_{i_1i_2\ldots i_d}\bfm{u}_{i_1i_2\ldots i_d})^{\T}\bSi_{i_1i_2\ldots i_d}^{-1}
(\by_{i_1i_2\ldots i_d} - \bF_{i_1i_2\ldots i_d}\bfm{u}_{i_1i_2\ldots i_d})+\lambda J(\mathcal{U})
 \end{equation}
for $\mathcal{U}\in\bbS$, where $\bbS$ is the parameter space for $\mathcal{U}$.

Based on the proposed method, $\widehat{\by}_{i_1i_2\ldots i_d}$ can be rewritten as $\widehat{\by}_{i_1i_2\ldots i_d}=\bW_{i_1i_2\ldots i_d}\bga$,
where $\bW_{i_1i_2\ldots i_d}=(\bX_{i_1i_2\ldots i_d1},\cdots,\bX_{i_1i_2\ldots i_dr},\bZ_{i_1i_2\ldots i_d1},\cdots,\bZ_{i_1i_2\ldots i_dm_{d+1}})$, $\bX_{i_1i_2\ldots i_dj}=u_{i_1i_2\ldots i_dj}\bB_{i_1i_2\ldots  i_dj}$,
$\bZ_{i_1i_2\ldots i_de}=u_{i_1i_2\ldots i_d(r+1)}\bA_{i_1i_2\ldots i_de}$,
in which
$\bB_{i_1i_2\ldots i_dj}=\{\bB_j(t)^{\T}\}_{t\in\bbT_{i_1i_2\ldots i_d}}\in\bbR^{|\bbT_{i_1i_2\ldots i_d}|\times M}$,
and
$\bA_{i_1i_2\ldots i_de}=\{I(t\in s_e)\bA_e(t)^{\T}\}_{t\in\bbT_{i_1i_2\ldots i_d}}\in\bbR^{|\bbT_{i_1i_2\ldots i_d}|\times M}$
for
$j=1,2,\ldots,r$,
$e=1,2,\ldots,\\ m_{d+1}$.
 By the approximation theory \citep{DeBoor2001}, there exists a constant $C>0$, the spline functions $\widetilde{h}_j(t)=\bal_{0j}^{\T}\bB_j(t)$ and $\widetilde{g}_e(t)=\bbe_{0e}^{\T}\bA_e(t)$ such that $\sup_{t\in\bbT}|h_j(t)-\widetilde{h}_j(t)|\le C a_N^{-\xi}$ and $\sup_{t\in\bbT}|g_e(t)-\widetilde{g}_e(t)|\le C a_N^{-\xi}$ for any $j=1,\dots,r$, $e=1,\ldots,m_{d+1}$. Denote $\bga_0=(\bal_0^{\T},\bbe_0^{\T})^{\T}$, and let $\lambda_{\min}\{\cdot\}$ and $\lambda_{\max}\{\cdot\}$ be the smallest and largest eigenvalues of any symmetric matrix, respectively.
We require the following regularity conditions to establish the asymptotic properties.

\begin{itemize}
\item[(C1)] The functions $h_j(\cdot)$ and $g_e(\cdot)$ are $\xi$th-order continuously differential for some $\xi\ge2$, all $j=1,\ldots,d$, and $e=1,\ldots,m_{d+1}$.
The density function of design points $t$ is absolutely continuous and bounded away from zero and infinity on a compact support $\bbT$.
\item[(C2)]
The knots sequences $\Upsilon_j$ and $\Gamma_e$ are quasi-uniform for $j=1,\ldots,d$ and $e=1,\ldots,m_{d+1}$; that is, there exists a constant $c>0$, such that
$$
\max_{j=1,\ldots,d}\frac{\max_{i=0,\ldots,a_N}(\nu_{ji+1}-\nu_{ji})}{\min_{i=0,\ldots,a_N}(\nu_{ji+1}-\nu_{ji})}\le c,\;\;{\rm{and}}\;\;
\max_{e=1,\ldots,m_{d+1}}\frac{\max_{i=0,\ldots,a_N}(\omega_{ei+1}-\omega_{ei})}{\min_{i=0,\ldots,a_N}(\omega_{ei+1}-\omega_{ei})}\le c.
$$

\item[(C3)] There exist positive constants $\sigma_1^2$ and $\sigma_2^2$ such that the covariance matrix $\bSi^0_{i_1i_2\ldots i_d}$ of random error $\bvarep_{i_1\ldots i_d}$ satisfies that
    $\sigma_1^2\le \lambda_{\min}\{\bSi^0_{i_1i_2\ldots i_d}\}\le\lambda_{\max}\{\bSi^0_{i_1i_2\ldots i_d}\}\le \sigma_2^2$.

\item[(C4)] There exist some positive constants $c_1$ and $c_2$ such that
$c_1\le \lambda_{\min}\{\bSi_{i_1i_2\ldots i_d}^{-1}\bSi^0_{i_1i_2\ldots i_d}\}\le\lambda_{\max}\{\bSi_{i_1i_2\ldots i_d}^{-1}\bSi^0_{i_1i_2\ldots i_d}\}\le c_2$.

\item[(C5)]

$T_{\max}=\max_{(i_1,\cdots, i_d)\in\Omega}\{|\bbT_{i_1\cdots i_d}|\}=o_p(N^{\tau})$,
$T_{\min}=\min_{(i_1,\cdots, i_d)\in\Omega}\{|\bbT_{i_1\cdots i_d}|\}=o_p(N^{\upsilon})$
for $0\le\tau/2<\upsilon\le\tau<1$,
and $\lambda=o_p(1)$.
\end{itemize}

Conditions (C1)-(C3) are standard in the polynomial spline framework. Similar conditions are also presented in \cite{huang2003} and \cite{claeskens2009}. In particular,
condition (C1) imposes a smoothness condition of trend functions and a mild condition on time density, and guarantees that the observation time points are randomly scattered.
Condition (C2) indicates that the adjacent distances among the knot sequence are comparable.
Condition (C3) implies that the eigenvalues of random errors are bounded.
Condition (C4) implies that the difference between the working covariance and true covariance matrices is bounded.
Condition (C5) implies that the number of the observed time points grows as the number of the observed components of the tensor increases, to ensure the convergence of the proposed
tensor factorization.
The following theorem establishes the convergence rate for the proposed tensor
factorization.

\begin{theorem}\label{thm1}
Under conditions (C1)-(C5), if the penalty function $J(\mathcal{U})$ has bounded first and second derivatives at true parameter $\mathcal{U}_0$, as $N\rightarrow\infty$,
on a $\delta$-ball centered at $\mathcal{U}_0$ for some $\delta>0$, there exists a minimizer $\widehat{\mathcal{U}}$ of (\ref{objU}) such that
$$
\sum_{(i_1,i_2,\cdots, i_d)\in\Omega}\|\bF_{i_1i_2\cdots i_d}(\widehat{\bu}_{i_1i_2\cdots i_d}-\bu_{0i_1i_2\cdots i_d})\|_2^2/N =O_p(N^{-1+2(\tau-\upsilon)}).
$$
\end{theorem}

Theorem \ref{thm1} provides the convergence rate of the proposed method given trend functions. When $\tau=\upsilon$, that is, $T_{\max}$ and $T_{\min}$ have the same order, the convergence rate of the estimator $\widehat{\mathcal{U}}$ reaches the optimal rate $N^{-1/2}$. Meanwhile, if the order of $T_{\max}$ is $\sqrt{N}$ faster than that of $T_{\min}$, that is, $\tau-\upsilon=0.5$, then $\widehat{\mathcal{U}}$ will not converge to the true $\mathcal{U}_0$. This implies that to guarantee consistency of the tensor factorization, one should collect sufficient observations even for the least popular user-item-context combinations.
In the following theorem, we establish the asymptotic property of the spline coefficient estimator.

\begin{theorem} \label{thm2}
Under conditions (C1)-(C5), if $\lim_{N\rightarrow\infty} a_N\log a_N/N=0$
and $\lim_{N\rightarrow\infty} a_N^{-\xi}N^{\tau}\\ =0$,
then for any vector $\bc$ whose components are not all zero,
the parametric estimator $\hbga$ by (\ref{objf3}) and (\ref{objf4}) satisfies
$$
\bc^{\T}(\hbga-\bga_0)var\{\bc^{\T}(\hbga-\bga_0)\}^{-1/2}\srf{L}{\rightarrow} N(0,1),
$$
where
$var\{\bc^{\T}(\hbga-\bga_0)\}=\bc^{\T}\bfm{\Psi}^{-1}\bfm{\Phi}\bfm{\Psi}^{-1}\bc=O_p(a_NN^{-1+\tau-2\upsilon})$,
$\bfm{\Psi}=\sum_{(i_1,\ldots, i_d)\in\Omega}\bW_{i_1\ldots i_d}^{\T}\bSi_{i_1\ldots i_d}^{-1}\cdot\\ \bW_{i_1\ldots i_d}$,
and
$\bfm{\Phi}=\sum_{(i_1,\ldots, i_d)\in\Omega}\bW_{i_1\ldots i_d}^{\T}\bSi_{i_1\ldots i_d}^{-1}\bSi^0_{i_1\ldots i_d}\bSi_{i_1\ldots i_d}^{-1}\bW_{i_1\ldots i_d}$.
\end{theorem}

Theorem \ref{thm2} establishes the asymptotic normality of the spline coefficient estimator.
The convergence rate of the spline coefficient estimator is $O_p(a_NN^{-1+\tau-2\upsilon})$.
If $T_{\max}$ and $T_{\min}$ have the same order, $var\{\bc^{\T}(\hbga-\bga_0)\}=O_p(a_N/N^{1+\upsilon})$, and similar results can be found in \cite{Huang2004Wu}. The asymptotic variance in Theorem \ref{thm2} depends on the working covariance matrix and the true covariance matrix.
When the working covariance matrices are equal to the true covariance matrices, the asymptotic variance of the proposed estimator reaches the minimum
 in the sense of Loewner order and the proposed estimator is asymptotic efficient.

More importantly, the result of Theorem \ref{thm2} is the key foundation for constructing prediction intervals. First, we derive the standard error for the spline parametric estimates given a fixed $\lambda$ using the sandwich covariance formula
$\widehat{\Cov}(\hbga)=(\widehat{\bfm{\Psi}}+\lambda\bI)^{-1}\widehat{\bfm{\Phi}}(\widehat{\bfm{\Psi}}+\lambda\bI)^{-1},$
where $\widehat{\bfm{\Psi}}=\sum_{(i_1,i_2,\ldots,i_d)\in\Omega}\hbW_{i_1i_2\ldots i_d}^{\T}\bSi_{i_1i_2\ldots i_d}^{-1}\hbW_{i_1i_2\ldots i_d}$,
$\widehat{\bfm{\Phi}}=\sum_{(i_1,i_2,\ldots,i_d)\in\Omega}\{\hbW_{i_1i_2\ldots i_d}^{\T}\bSi_{i_1i_2\ldots i_d}^{-1}(\by_{i_1i_2\ldots i_d}-\hbW_{i_1i_2\ldots i_d}\hbga)\}^{\otimes2}$, $\otimes$ operation is the vector operation $\bfm{a}^{\otimes2}=\bfm{a}\bfm{a}^{\T}$, and
$\bI$ is an identity matrix.
Since $\hy_{i_1i_2\ldots i_d}(t)=\hbw_{i_1i_2\ldots i_dt}^{\T}\hbga$, and $\hbw_{i_1i_2\ldots i_dt}$ is the $t$th column of estimator $\hbW_{i_1i_2\ldots i_d}^{\T}$,
a $100(1-\sigma)\%$ prediction interval \citep{Chris1993} of $\hy_{i_1i_2\ldots i_d}(t)$ is
\begin{equation}\label{confidence}
\hy_{i_1i_2\ldots i_d}(t)\pm\phi_{\sigma/2}\sqrt{\var\{e_{i_1i_2\ldots i_d}(t)\}},
\end{equation}
where $\phi_{\sigma/2}$ is the $100(1-\sigma)$th percentile of the standard normal distribution, and the $\var\{e_{i_1i_2\ldots i_d}(t)\}$ is the variance of the prediction error and can be estimated as:
\begin{equation}\label{varprediction}
\widehat{\var}\{e_{i_1i_2\ldots i_d}(t)\}=\hbw_{i_1i_2\ldots i_dt}^{\T}\widehat{\Cov}(\hbga)\hbw_{i_1i_2\ldots i_dt} + \widehat{\var}\{\varepsilon_{i_1i_2\ldots i_d}(t)\}.
\end{equation}
The first term in equation (\ref{varprediction}) is due to estimation error, and the second term can be estimated by the mean squared error on training data.

\section{Simulation Studies}

In this section, we perform simulation studies to compare the proposed method (DTRS) with two competing methods, including Bayesian probabilistic tensor factorization \citep[BPTF,][]{xiong2010temporal} and the recommendation engine of multilayers \citep[REM,][]{bi2018}. We assess forecasting performance via examining the root mean square error (RMSE) and the mean absolute error (MAE), where the RMSE is defined as $[\sum_{(i_1,\ldots,i_d,t)\in\Gamma}\{y_{i_1\ldots i_d}(t)-\hy_{i_1\ldots i_d}(t)\}^2/|\Gamma|]^{1/2}$, the MAE is defined as $\sum_{(i_1, \ldots, i_d,t)\in\Gamma}|y_{i_1\ldots i_d}(t)-\hy_{i_1\ldots i_d}(t)|/|\Gamma|$, and $\Gamma$ is the set of indices and times of observed data.  Moreover, we evaluate the coverage probability of the prediction interval estimated by the proposed method with $95\%$ nominal coverage probability (PICP) .

In the simulation, we consider a third-order tensor function of time with user, context and item modes. We set the numbers of users, contexts and items to $n_1=100$, $n_2=9$, and $n_3=100$, respectively. We assume that users, contexts, items and time points are from $m_1=10$, $m_2=3$, $m_3=10$ and $m_4=4$ subgroups, respectively. Users, contexts, items and time points are evenly assigned to each subgroup.
The number of latent factors is set as $r=3$. We generate tensor functions at time points $t\sim U(0,1)$ by generating its components as  $y_{i_1i_2i_3}(t)=\sum^r_{j=1}h_j(t)p_{i_1j}^1p_{i_2j}^2p_{i_3j}^3+g(t)q_{i_1}^1q_{i_2}^2q_{i_3}^3+\varepsilon_{i_1i_2i_3}(t)$
for $i_k=1,\ldots,n_k$, $k=1,2,3$, where the latent factors $\bp^k_{i_k}\sim N(0,\I_r)$, trend functions $h_1(t)=\sin(0.3\pi t)$, $h_2(t)=8t(1-t)-1$ and $h_3(t)=\cos(0.2\pi t)+1$. To distinguish different subgroups, we set the subgroup factors as a simple sequence, where $\bq^1_{(e_1)}=-1+0.4e_1$, $\bq^2_{(e_2)}=-1.2+0.6e_2$ and $\bq^3_{(e_3)}=-0.4+0.2e_3$ for $e_k=1,\ldots,m_k$ and $k=1,2,3$. The function $g(t)=\sum_{e=1}^{m_{4}}g_e(t)I(t\in s_e)$, where $g_1(t)=2t-1$, $g_2(t)=8(t-0.5)^3$, $g_3(t)=\sin(0.1\pi t)+\cos(\pi t)$, and $g_4(t)=-5\exp(t)+10$.
The error $\bvarep_{i_1i_2i_3}=(\varepsilon_{i_1i_2i_3}(t_1),\ldots,\varepsilon_{i_1i_2i_3}(t_T))^{\T}$ follows a multivariate normal distribution with mean 0 and a common marginal variance 1, and the correlation structure is either independence or AR-1 with correlation $\rho=0.85$.

In each simulation, we consider the number of time points as $T=T_1+T_2$, where the tensor data in the first $T_1=12$ time points are set as the training data, and the tensor data in the last $T_2$ time points are used as the testing data. For evaluating the forecasting performance at future time points, we consider $T_2=8$ or $12$.
Considering the missing case, we generate $n_1n_2n_3T(1-\pi_m)$ components out of the tensor functions, where $\pi_m$ is the missing percentage and set as $80\%$. Furthermore, we use $\pi_{cs}=30\%$ to represent the proportion of new items in the testing data unavailable from the training set.
To illustrate the effect of incorporating intra-cluster correlation on estimation efficiency, we compare the estimation efficiency of the proposed methods using different working correlation structures: independent or AR-1, denoted as DTRSin and DTRSar, respectively.

According to \cite{xiong2010temporal} and \cite{bi2018}, BPTF and REM methods model fourth-order tensor with user, context, item and time modes. For all methods, we assume that the subgroup structure and the number of latent factors are known.
For REM and the proposed methods, the tuning parameter $\lambda$ is pre-selected from grid points ranging from 0 to 20.
The validation set is the data from the last four time points of the training set.
For BPTF, we keep the remaining parameters by their default choices.
All methods are replicated by 100 simulation runs.

Table \ref{simulationtable1} provides the estimation results of all methods.
We observe that the proposed method has better performance when the working correlation structure is the same as the true correlation structure. When the true correlation structure is independence, the DTRSin has smaller RMSE and MAE than the DTRSar, with more than $2.17\%$ improvement. Similarly, when the true correlation structure is AR-1, the DTRSar outperforms the DTRSin. Moreover, the PICPs of the DTRS method are close to $0.95$, which implies that the proposed method provides accurate prediction intervals for estimators. For the performance of forecasting time points further away, we observe that the DTRSin, DTRSar and REM methods are relatively robust against time. However, the RMSE and MAE of the REM are larger than the RMSE and MAE of the DTRS method, and the BPTF performs worse on forecasting time points further away. Specifically, the DTRS method performs the best across all settings. For the DTRS method, the relative increasing ratios of RMSEs when $T_2=8$ to those when $T_2=12$ are less than $6.9\%$, and the corresponding MAEs are at most $4.0\%$. However, for the BPTF method, the relative increasing ratios of RMSEs for the two time points are more than $9.5\%$, and the corresponding MAEs are at least $8.1\%$. The DTRS method improves on the RMSE and MAE of the BPTF by more than $65\%$, and improves the RMSE and MAE of the REM by more than $44\%$. This indicates that the proposed method can obtain more accurate forecasting compared to the BPTF and REM.


\begin{table}  [H]
\begin{center}
\caption{Average RMSE and MAE of all approaches. The PICP is average coverage probability of $95\%$ prediction interval. The RMSE, MAE and PICP are provided with standard error based on 100 simulations in each parenthesis. } \label{simulationtable1}  
\vspace{0.2cm}
\begin{tabular}{cc|cc|ccccccccccccc}  \hline  \hline
\MC{2}{c}{True structure:}& \MC{2}{c}{Independent}            &  \MC{2}{c}{AR}  \\ \hline
Method&      &  $T_2=8$       & $T_2=12$      & $T_2=8$       & $T_2=12$  \\  \hline
DTRSin& RMSE &  \textbf{1.570}(0.196)  & \textbf{1.660}(0.389)  & 1.597(0.192)  & 1.707(0.524)\\
      & MAE  &  \textbf{1.092}(0.091)  & \textbf{1.132}(0.160)  & 1.115(0.091)  & 1.160(0.208)\\
      & PICP &  \textbf{0.949}(0.015)  & \textbf{0.953}(0.017)  &  0.946(0.017) & 0.952(0.018)\\
DTRSar& RMSE &  1.625(0.244)  & 1.696(0.286)  & \textbf{1.576}(0.190)  & \textbf{1.632}(0.200)\\
      & MAE  &  1.133(0.118)  & 1.170(0.159)  &  \textbf{1.099}(0.085) & \textbf{1.130}(0.102) \\
      & PICP &  0.943(0.019)  & 0.947(0.021)  & \textbf{0.947}(0.015)  & \textbf{0.949}(0.018)\\
BPTF  & RMSE &  2.675(0.742)  & 2.930(0.965)  & 2.724(0.863)  & 3.181(1.148) \\
      & MAE  &  1.810(0.427)  & 1.958(0.547)  & 1.826(0.495)  & 2.104(0.654) \\
REM   & RMSE &  2.502(0.322)  & 2.494(0.307)  & 2.498(0.304)  & 2.494(0.305) \\
      & MAE  &  1.654(0.178)  & 1.640(0.170)  & 1.650(0.166)  & 1.643(0.172) \\
                \hline
\hline
\end{tabular}
\end{center}
\end{table}

\begin{figure} [H]
\centering
\scalebox{0.45}[0.45]{\includegraphics{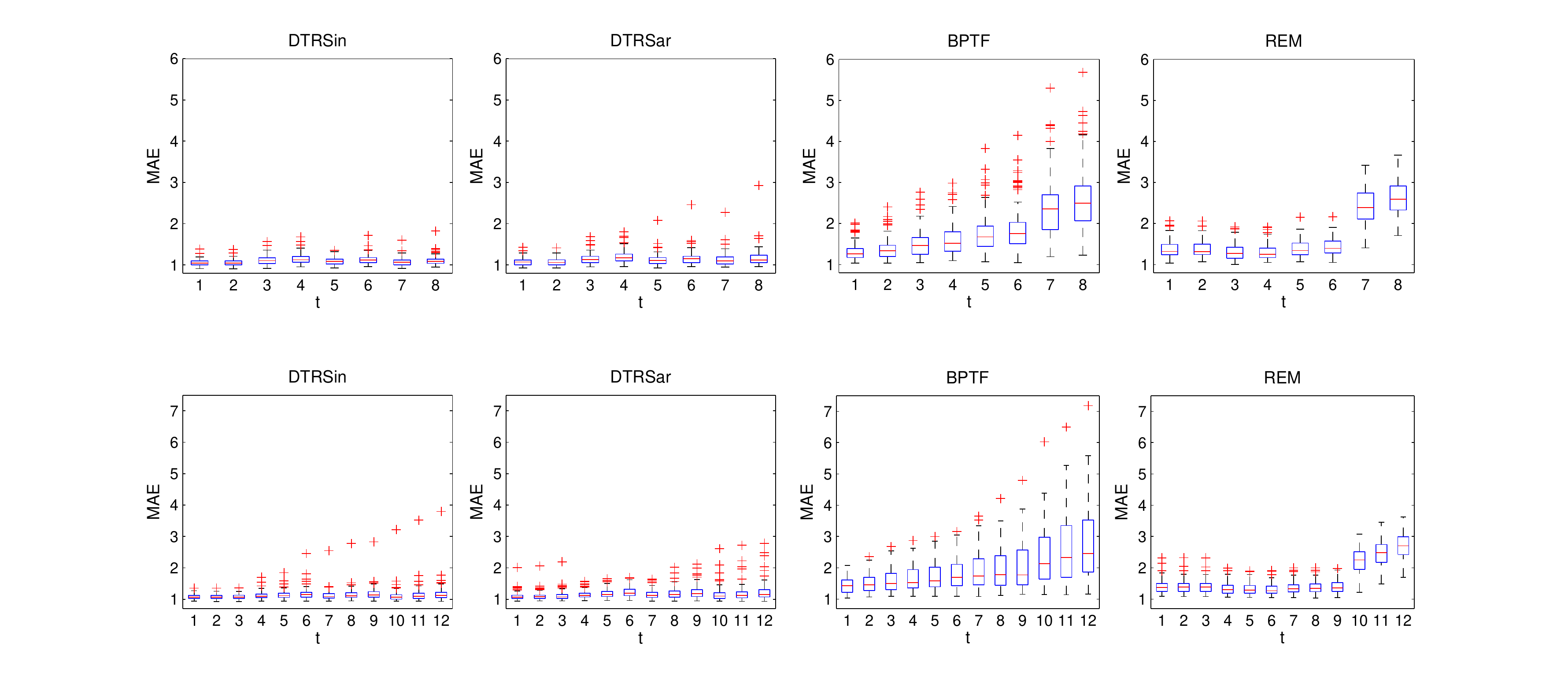}} \vspace{-12mm}
\caption{\footnotesize{Box plots of the MAE for forecasting values with 8 and 12 time points and true independent correlation.
}}
\label{boxplot_independent8}
\end{figure}

\begin{figure} [H]
\centering
\scalebox{0.45}[0.45]{\includegraphics{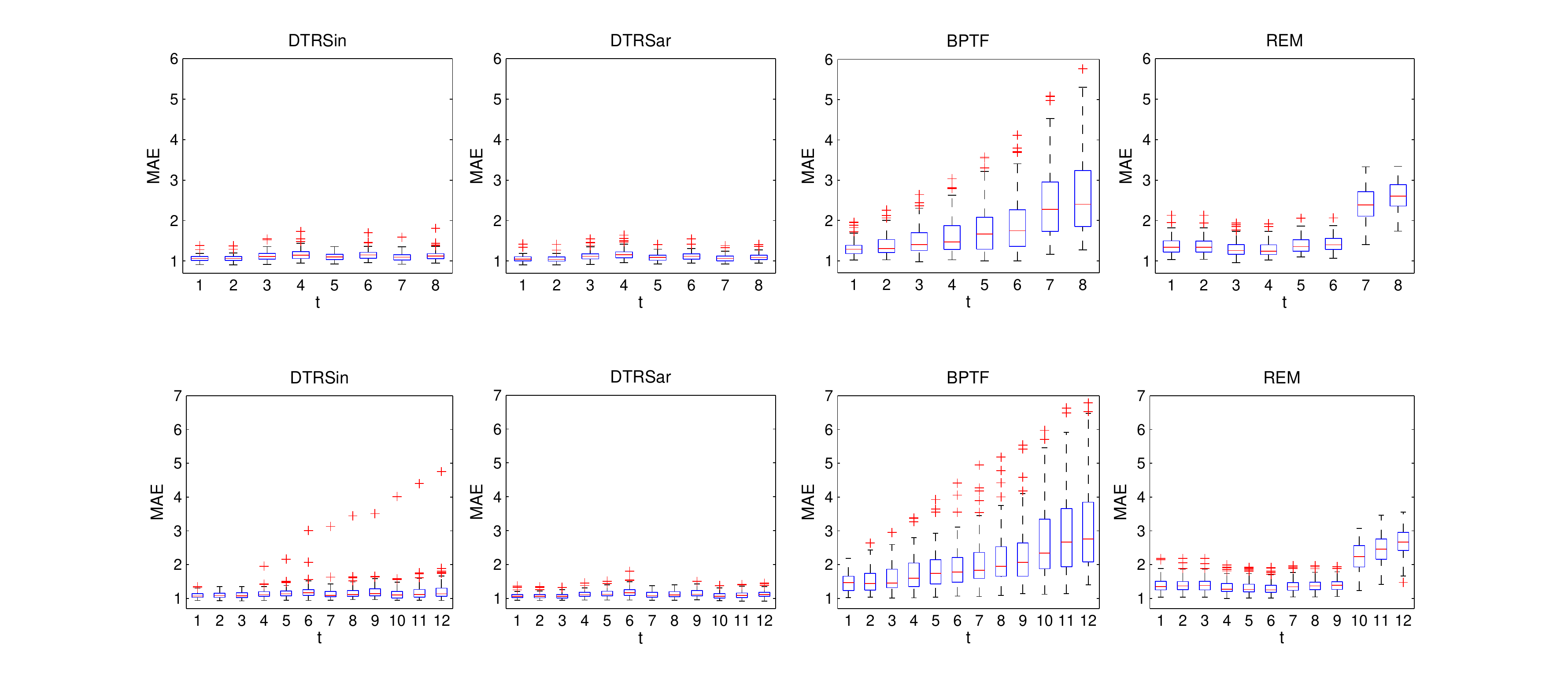}}  \vspace{-12mm} 
\caption{\footnotesize{Box plots of the MAE for forecasting values with 8 and 12 time points and true AR-1 correlation.
}}
\label{boxplot_AR8}
\end{figure}

To illustrate the specific performance for forecasting at each time point, we calculate the MAE at each time point and provide box plots for the MAE in Figures \ref{boxplot_independent8}-\ref{boxplot_AR8}. We observe that the performance of the proposed method is relatively robust against time in all settings. The MAEs of BPTF and REM increase for time points further away. The MAEs of both DTRS methods at all time points are lower than those for the other two methods. This indicates that the proposed method outperforms other methods with respect to forecasting at later time points.

\section{Empirical Examples for IRI Marketing Data}

In this section, we focus on sales data from drug stores from the IRI Marketing Data
\citep{Bronnenberg2008} to illustrate the performance of the proposed method.
The original IRI data is an immense collection of consumer panel data and store sales at grocery stores, drug stores and mass-market stores over the years 2001-2011. The store sales data contain weekly product sales volumes, pricing, and promotion data for all items from 31 product categories sold in 50 U.S. markets.
These markets are geographic units defined typically as an agglomeration of counties, usually covering a major metropolitan areas (e.g., Chicago, IL) but sometimes covering just part of a region (e.g., New England).
A detailed description of an early version of the data is available in
\cite{Bronnenberg2008}.

To illustrate the proposed method, we choose sales data at drug stores collected from 2001 to 2011, where there are sales volume records,  recorded times,  promotion strategies, 43,631 product IDs, and 471 drug store IDs. These drug
stores are from 50 markets across the United States. The products include items sold from these stores during the 11-year
period, and are from 31 product categories, including hot dogs, household cleaners, margarine/butter blends, mayonnaise, milk, coffee, cigarettes, photography supplies, paper towels, frozen pizza, toilet tissue,
yogurt, beer/ale/alcoholic cider, blades, cold cereal, carbonated beverages,
diapers, deodorant, facial tissue, frozen dinners/entrees, laundry detergent,
peanut butter, razors, mustard and ketchup, sugar substitutes, spaghetti/Italian sauce, soup, shampoo, salty snacks, toothpaste, and toothbrush. Moreover, various advertising and promotions strategies are imposed on these products to attract consumers. The promotions strategies have 30 types which are combinations
of 5 advertisement features, 3 types of merchandise display, and
an indicator on whether the product has a price reduction of more than 5\%.

The goal of our study is to predict the future sales volumes of each product from each store given each promotion strategy based on historical sales data. Through this prediction procedure, we are able to estimate future purchases, evaluate the influence of promotion strategy for product sales, and potentially recommend the most profitable products to store managers, so the company can make wiser decisions on marketing strategies and inventory planning. For considering the trend of product sales, we aggregate the weekly data into monthly data
according to the record time information
so that the data contain more than 79.2 million sales records for 132 months from the beginning of 2001 to the end of 2011. For the proposed method, we classify stores, products, observed time points and promotion strategies into subgroups based on their markets, product categories, month of the year and whether a price reduction is applied, respectively.

\begin{table}  [H]
\begin{center}
\caption{Summary statistics of the monthly IRI marketing data.}  \label{realdatasummary}
\begin{tabular}{ccccccccccccccccc}  \hline  \hline
             & The number of types & The number of subgroups  \\ \hline
Store        & 471             & 50   \\
Promotion    & 30              & 2  \\
Product      & 43,631          & 31   \\
Month        & 132             & 12    \\
Sales record  & 79,243,289        \\
\hline
\end{tabular}
\end{center}
\end{table}

\begin{figure} [H]
\centering
\scalebox{0.4}[0.3]{\includegraphics{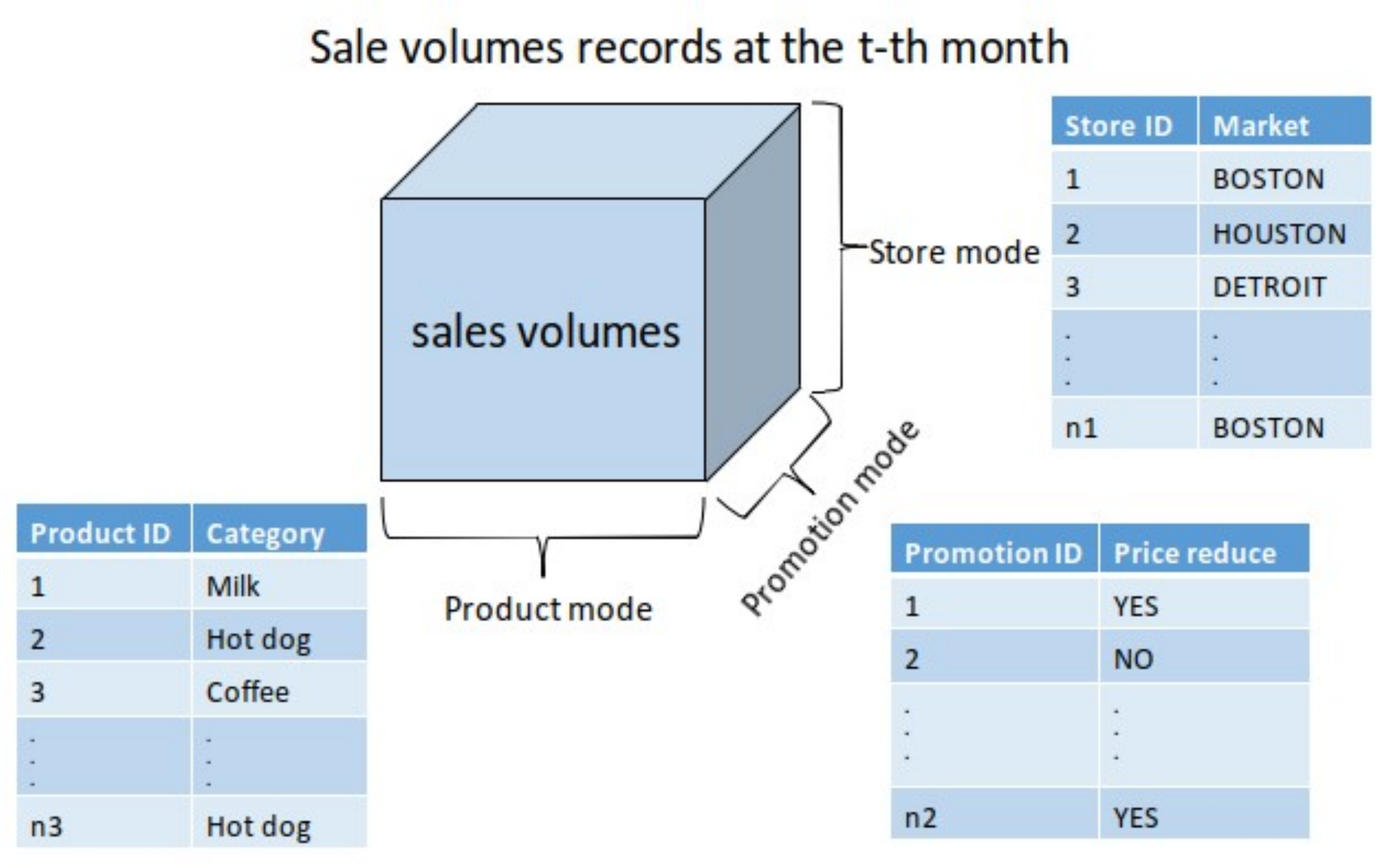}} 
\caption{\footnotesize{An illustration for monthly sale tensors at the $t$th month.}}
\label{saletensor}
\end{figure}

Table \ref{realdatasummary} shows the summary statistics of the data. According to the proposed method, the data can be reframed into monthly third-order tensors by store, product and promotion. Figure \ref{saletensor} provides an illustration of the reframed sale records to clarify the proposed method. According to the given structure of monthly third-order tensors, the total number of sale records could be up to $471\times30\times43631\times132\approx8.1$ billion. Although the observed records are more than 79.2 million, there are still a large number of store-promotion-product-month combinations which are associated with unknown sales volumes.
The sales data have a 99.9\% missing rate and are highly sparse, which renders a particular challenge for recommender systems and forecasting.

For comparison, we implement and report the performances of two competing methods and the proposed methods with different working correlation matrices as in Section 5. For all methods, we select the number of latent factors $r$ ranging from 3 to 30.
For the REM method and the proposed methods, we select a tuning parameter $\lambda$ from 1 to 29.
For the BPTF, we use the default values of the remaining parameters.
For selecting the above  parameters,
we set the data from the beginning of 2001 to the end of 2009 (i.e., the first 108 months) as the training set and the data from the beginning of 2010 to the end of 2010 as the validation set, and then tune these parameters through minimizing the root mean square error on the validation set.
Then we use the data from the entire year of 2011 as the testing set and predict the sales volumes based on historical sales data from 2001 to 2010.
There exist $2502$ new products in the testing data unavailable from the training set.

\begin{figure} [H]
\centering
\scalebox{0.4}[0.25]{\includegraphics{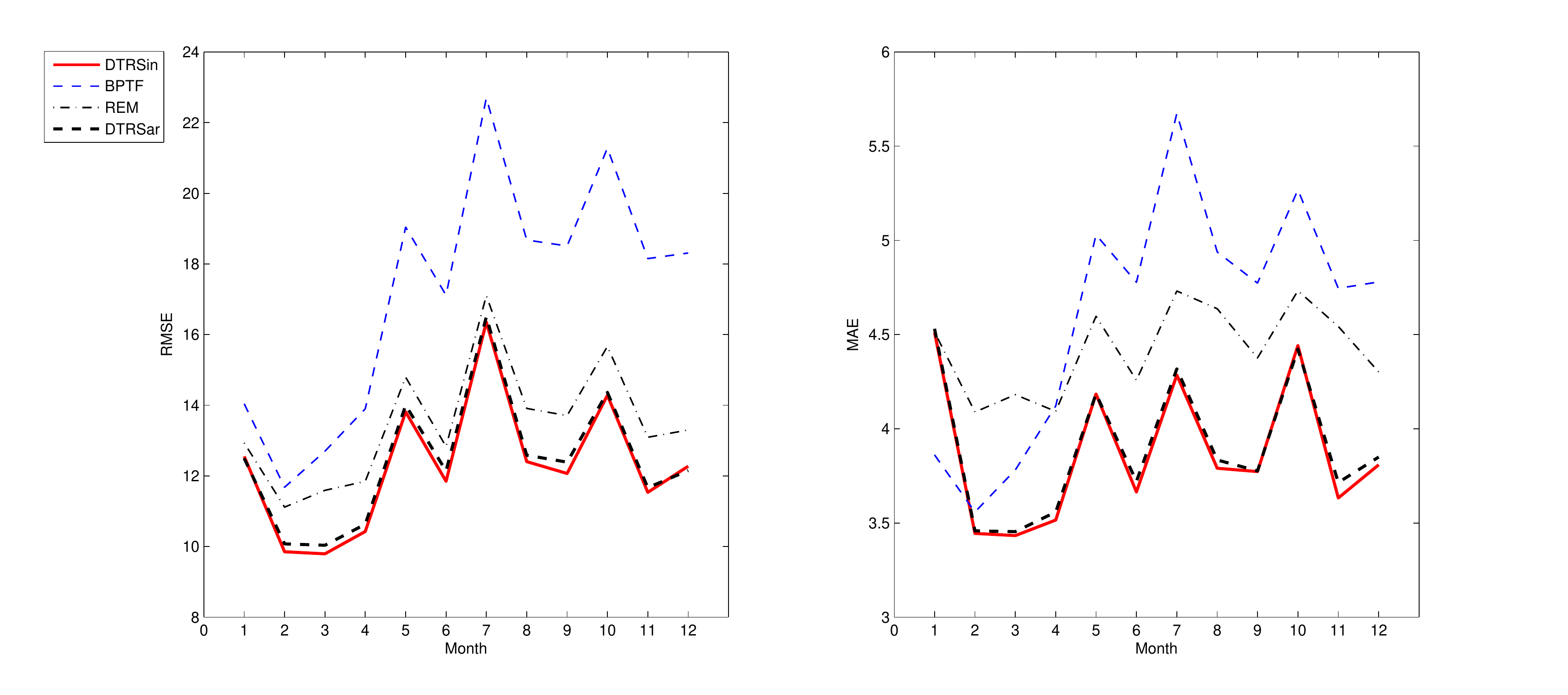}} \vspace{-5mm}
\caption{\footnotesize{Monthly RMSE and MAE of the forecasting sale volumes on each month in 2011. The solid lines indicate the results of DTRSin, the thick black dash lines indicate the results of DTRSar, the blue dash lines indicate the results of BPTF, and the dash-dotted lines indicate the results of REM.}}
\label{RMSEMAE}
\end{figure}

Table \ref{realdata} shows forecasting results produced by each method. The RRMSE and RMAE show the relative improvement ratios of the DTRSin method over the other methods in terms of the RMSE and MAE.
From Table \ref{realdata}, we observe that
the DTRSin method has the lowest RMSE and MAE and improves on the RMSE and MAE of DTRSar by $6.1\%$ and $4.6\%$, respectively.
This implies that the independent working correlation might be the most appropriate for this data, likely because the high missing rate weakens
the intra-cluster correlation. The DTRSin improves on the RMSE and MAE of BPTF by the largest percentages, that is, $32.5\%$ and $14.8\%$, respectively, and improves on the RMSE and MAE of REM by $9.4\%$ and $13.7\%$, respectively. This shows that the DTRSin method outperforms the BPTF and REM methods in predictions.
For illustrating the performance in each month, we show the average monthly RMSE and MAEs in Figure \ref{RMSEMAE},
where the solid lines indicate the results of DTRSin, the thick black dash lines indicate the results of DTRSar, the blue dash lines indicate the results of BPTF, and the dash-dotted lines indicate the results of REM.
Figure \ref{RMSEMAE} shows that the monthly RMSE and MAE of the DTRSin are lower than those of the BPTF and REM methods, and the DTRSar has similar performance as the DTRSin. The RMSE of BPTF has an upward trend in general over time, but the RMSEs of the DTRSin and DTRSar only fluctuate around 13. This implies that the proposed method has relatively more stable performance on forecasting future sale volumes when the predicted months are further away.

\begin{figure} [H]
\centering
\scalebox{0.45}[0.42]{\includegraphics{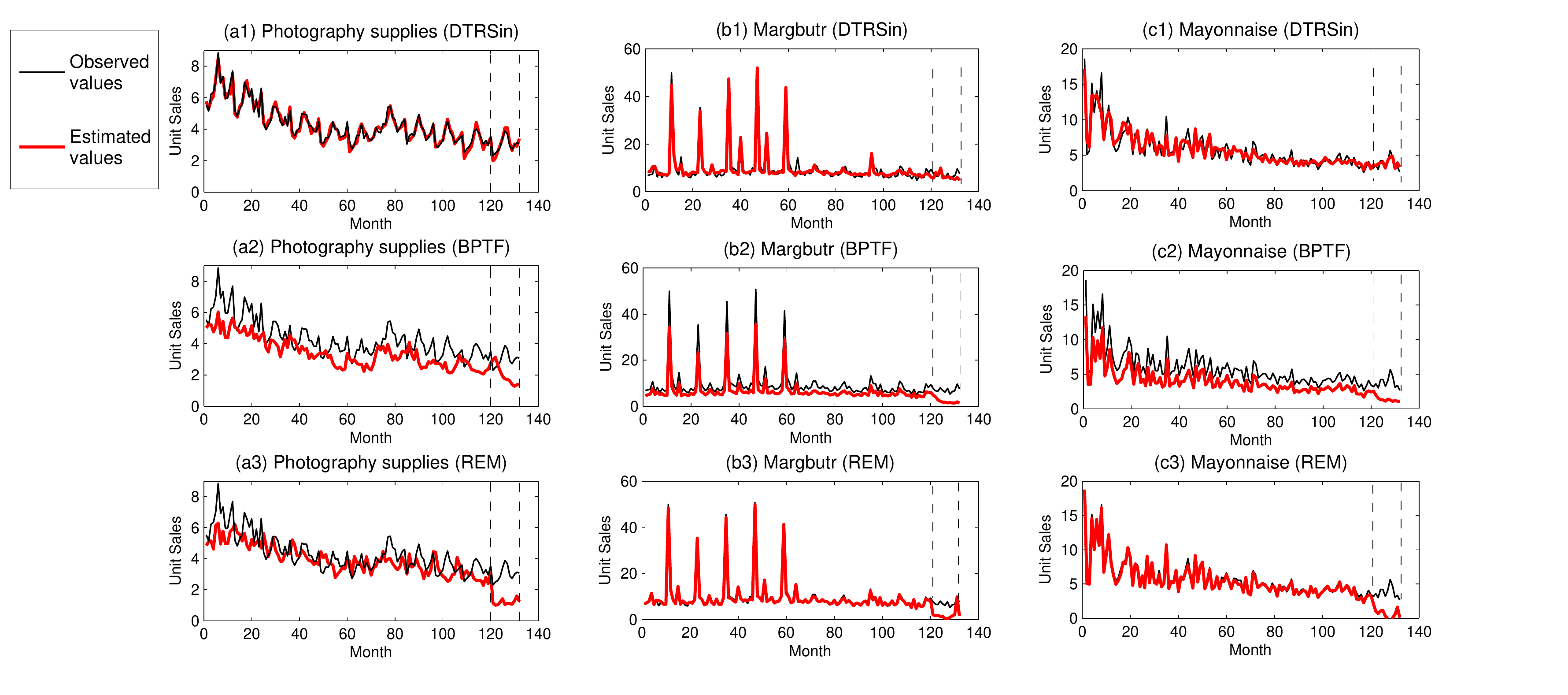}} \vspace{-12mm}
\caption{\footnotesize{Average unit sales for three categories of products over time: photography supplies, margarine/butter blends, and mayonnaise. The black lines indicate the observed values, and the red lines indicate the estimated values. The time intervals from the 121th month to the 132th month show the forecasting performance. The (a1)-(c1) show the results of DTRSin, the (a2)-(c2) show the results of BPTF, and the (a3)-(c3) show the results of REM.
}}
\label{forecast}
\end{figure}

Moreover, we show average sales volumes for three arbitrary categories of products over 132 months in Figure \ref{forecast}, and the prediction interval estimated by the proposed method in Figure \ref{PI}, while the corresponding figures for the rest of the product categories can be found in the Supplementary Material. In Figure \ref{forecast}, the black lines indicate the observed values, and the red lines indicate the estimated values. The time intervals from the 121st month to the 132nd month show the forecasting performance. The DTRSar has performance similar to the DTRSin, and is therefore not provided here. We notice that the DTRSin can estimate forecasting more accurately. Although the REM can estimate sales volumes sufficiently well on the training set, the forecasting on the testing data has relatively larger biases, while the BPTF has poor performance on both the training and testing sets.
Figure \ref{PI} provides pointwise prediction intervals for average sale volumes estimators under $95\%$ and $50\%$ nominal coverage probabilities.

\begin{table}  [H]
\begin{center}
\caption{ The
RMSE and MAE of the forecasting sale volumes in 2011 by four methods.
The RRMSE and RMAE show the relative improvement ratios of the DTRSin method over others in terms of the RMSE and the MAE.
}  \label{realdata}
\vspace{0.2cm}
\begin{tabular}{ccccccccccccccccc}  \hline  \hline
Method   & RMSE             & RRMSE   & MAE    & RMAE       \\ \hline
DTRSin     & \textbf{12.504}  & --      & \textbf{3.897}  & --         \\
DTRSar    & 13.266  & 6.1\%   & 4.075  & 4.6\% \\
BPTF     & 16.571           & 32.5\%  & 4.474  & 14.8\%     \\
REM      & 13.679           & 9.4\%   & 4.432  & 13.7\%      \\
\hline
\end{tabular}
\end{center}
\end{table}

\begin{figure} [H]
\centering
\scalebox{0.46}[0.35]{\includegraphics{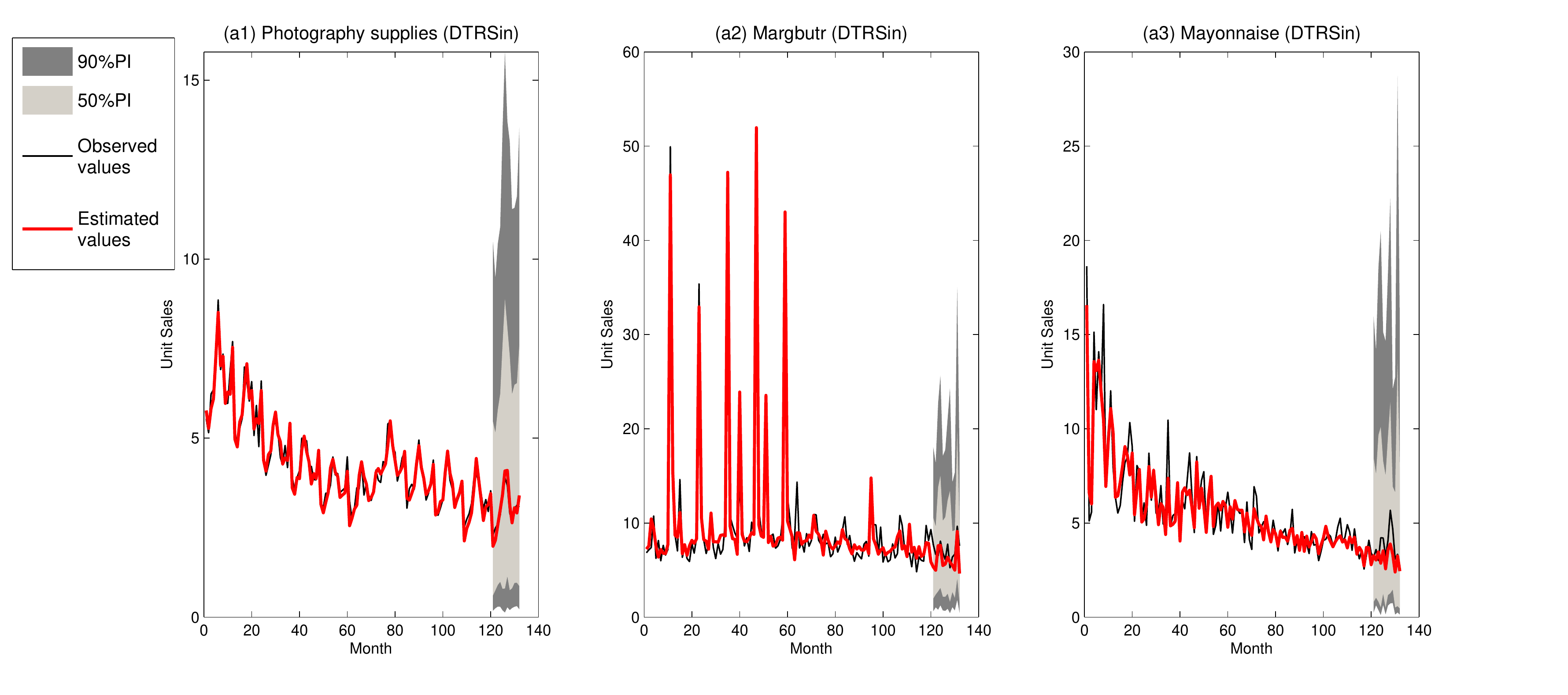}} \vspace{-12mm}
\caption{\footnotesize{Average unit sales and prediction intervals for three categories of products over time: photography supplies, margarine/butter blends, and mayonnaise.
}}
\label{PI}
\end{figure}

\section{Discussion}

In this article, we propose a new dynamic tensor recommender system which incorporates time information through a tensor-valued function. A unique contribution of the proposed method is that it can effectively forecast future recommendations at irregular time points. Technically, the proposed method builds a time-value tensor decomposition model and borrows group information from existing time points of the same group for higher forecasting accuracy. Moreover, the proposed method
 utilizes the polynomial spline method and the weighted least squared method to incorporate time-dependency and intra-cluster correlation into the DRS.
In addition, the proposed method is able to provide pointwise prediction intervals based on the established asymptotic property, while existing recommender systems are not equipped with prediction intervals. In theory, we demonstrate that the proposed decomposition achieves asymptotic consistency on prediction and the spline coefficient estimators have asymptotic normality. The proposed method shows numerical advantages compared to existing methods. In real example analysis for the IRI marketing data, the proposed method achieves better performance on forecasting than competitive approaches.

\vskip 14pt
\noindent {\large\bf Supplementary Material}

The online Supplementary Material contains the other results of real data analysis and all technical conditions and proofs.
\par

\vskip 14pt
\noindent {\large\bf Acknowledgments}
We would like to acknowledge support for this project
from the National Science Foundation Grants (DMS1821198 and DMS1613190), China Postdoctoral Science Foundation Grant (2017M623077),
and National Natural Science Foundation of P.R. China (11601471, 11731011, 11871420).
We would like to thank IRI for making the data available. All estimates and analysis in
this paper based on data provided by IRI are by the authors and not by the IRI.
\par


\vskip 0.2in
\bibliographystyle{apalike}
\bibliography{reference}

\end{document}